\documentstyle[osa,preprint]{revtex}
\begin{document}

\title{RESISTANCE OF LAYERED SUPERCLEAN SUPERCONDUCTORS AT LOW TEMPERATURES}

\author{\bf A.I. Larkin \cite{PA} and Yu.N. Ovchinnikov \cite{PA}}

\address{Theoretical Physical Institute, University of Minnesota,\\
116 Church St. SE Minneapolis, MN 55455}

\maketitle

\begin{abstract}

The low energy excitation spectrum is found for a  
layered superconductor vortex with  a small number of  
impurities inside the vortex core.  All
levels are found to be correlated. This leads to the strong enhancement of
conductivity in superclean layered superconductors.
\end{abstract}

\section{Introduction}

In superconductors with weak pinning, the $I-V$ characteristic displays
anomalous properties \cite{1,2}. Some of them are very difficult to explain
in the framework of the quasiclassical approach. In the case of the  quasiclassical
approach there are three limiting cases, determined by values of the three
parameters: the size of the gap $\Delta$ in the single particle excitation
spectrum, the level spacing inside  the  vortex core $\omega_0\sim\Delta^2/\varepsilon_F$
and the electron mean collision time $\tau_{tr}$. The three limiting
cases are: "dirty" limit, for  $\tau_{tr}\Delta\ll 1$; "clean" limit  for 
$\Delta\gg \tau^{-1}_{tr}\gg \omega_0$, and superclean limit when the  
condition $\omega_0\tau_{tr}\gg 1$ is fulfilled.

In the ``dirty'' limit at zero temperature the calculation of Gor'kov
and Kopnin \cite{3} confirms the  qualitative picture of vortex motion of
Bardeen and Stephen \cite{4}. In accordance with the  picture of Bardeen, Stephen
the vortex core is in a ``normal'' state. Bardeen, Sherman \cite{5}  and
Larkin, Ovchinnikov \cite{6} derived  conductivity in a mixed state for low
temperatures and small magnetic field in the case of  moderately clean
superconductors. In this case compared  to the previous picture a logarithmic large
factor arises  in conductivity. This factor is related to   shrinkage of the
vortex core at low temperatures $T\ll T_c$ \cite{7}. The Hall component
of conductivity was found for moderately clean superconductors by
Kopnin and Lopatin \cite{8}.

The superclean case was studied in the Kopnin, Kravtsov paper \cite{9}.
It was found, that the level spacing $\omega_0$ inside a vortex core
plays  the same role as the cyclotron frequency $\omega_c=eH/mc$ in
a normal metal. It was also found, that in the superclean limit the Hall
component of the conductivity tensor is the largest one
$\sigma\simeq en_e/B$
(here $n_e$  is the electron density in the conduction band
, $B$ is the magnitude of the  magnetic field). The dissipative part of the
conductivity tensor is smaller by the parameter
$(\omega_0 \tau)^{-1}$.
Hence the dissipative part of the resistance tensor is the same as
in moderately clean superconductors. 

The quasiclassical approach
is probably violated in the two dimensional case (in layered superconductors),
because the excitation spectrum in the  vortex core is then discrete.

Guinea, Pogorelov \cite{10_1} consider the dissipation in the
vortex state as a result of  transitions between unperturbed levels 
induced by "moving" impurities.
Such a perturbation theory approach is valid 
only in the high velocity limit $v \gg v_F (\Delta/\epsilon_F)^2$.

Feigel'man, Skvorzov \cite{10} consider  energy dissipation during the 
vortex motion as a  result of Landau-Zener transitions between
levels. They suppose, that the level distribution inside the
vortex core obeys  Wigner-Dyson statistics subject to  some
corrections, related  to specifics of superconductivity \cite{11}.
Such a treatment  can probably be used in dirty and moderately
clean superconductors. This method,  although  differing from quasiclassical
approach, nevertheless  gives for the essential range of electrical fields the
same expression of conductivity as the quasiclassical approach.

In this paper we consider the  superclean limit. We find 
that in this region a  new mechanism of dissipation arises. In the
superclean limit no more than one impurity can be found  at  distances of order of the correlation length
$\xi=v_F/\Delta$ from the vortex center 
. It will be shown  that in such a case a  statistical
description of level positions is impossible. If an impurity is
placed at a distance of order of $\xi$ from the vortex center and
 is  weak (Born parameter is small), then the shift of levels is
also small. It is important also, that levels with even and
odd orbital momentum are shifted in opposite directions.
The level shift increases as impurity comes closer to the vortex center.
At some distance from the  impurity to the  vortex center levels
practically cross. It is very important, that all
levels with energy $|\varepsilon|\ll \Delta$  cross
simultaneously.

If we neglect the  weak (of order of $\omega_0/p_F\xi$)
repulsion of levels in this region, then positions of levels
as a  function of the  distance from the vortex core to the impurity  form two
families of crossing straight lines. Outside of the dangerous level crossing region 
these lines are practically horizontal (see fig.1). The size
of the dangerous zone, where the level lines can be considered as
crossing, depends on the Landau-Zener parameter and hence
on  the  vortex velocity. For the  vortex velocity $V$ in the range
$\omega_0\gg p_FV\gg \omega_0(\Delta/\varepsilon_F)$ the distribution
function of excitations inside the core of the  vortex does not  change
until the impurity comes into the dangerous zone. But  it changes
essentially when impurity goes  through the dangerous zone. Excitations,
that arise when impurity goes through this zone, determine the
value of the dissipative part of conductivity. Such  mechanism of
dissipation is essential for the magnitude  of the electrical
field $E$ lying in the range
\begin{displaymath}
B\frac{v_F}{c} \left (\frac{\Delta}{\varepsilon_F} \right)^2
\gg E \gg B \frac{v_F}{c} (\Delta/\varepsilon_F)^3
\end{displaymath}
Here $c$ is the velocity of light. As we prove below, this
mechanism of dissipation leads to the dissipative part of
conductivity $\sigma_{xx}$ being equal to
\begin{equation}
\sigma_{xx}\sim\frac{ecn_e(p_Fa_0)}{B(\omega_0\tau_{tr})}
\end{equation}
where $a_0$  is the distance from  the "dangerous" region to  the vortex center
$a_0\sim\theta\xi$. $\theta$  is the Born parameter which is
equal to the phase shift of an  electron scattering off the
impurity. Usually  its value is of order of one $\theta\sim 1$.
Hence parameter $(p_Fa_0)$ is much larger than one and
$\sigma_{xx}$ essentially exceeds the value, obtained in the
framework of the quasiclassical approximation. In the range
$p_Fa\gg \omega_0\tau_{tr}\gg 1$ the Hall angle is small.
Nevertheless the dependence of $\sigma_{xx}$ on the  
scattering time $\tau_{tr}$ is the same as at  large values
of the Hall angle.

\section{The low energy excitations spectrum for an impurity at the distance $a$
from the vortex center in the range
$a\gg \xi(\Delta/\varepsilon_F)^{1/2}$}

The excitations spectrum $E$ in the vortex state can be found as
a solution of the eigenvalue problem for the system of equations \cite{12,13}
\begin{equation}
\left (
\begin{array}{ll}
-\frac{1}{2m}
\frac{\partial^2}{\partial {\bf r}^2}-
\mu+V({\bf r})-E; & \Delta({\bf r}) \\
\Delta({\bf r})^*; & \frac{1}{2m}
\frac{\partial^2}{\partial {\bf r}^2}+\mu-V({\bf r})-E
\end{array} \right)
\left (
\begin{array}{c}
f_1 \\
f_2
\end{array} \right) =0
\end{equation}
where $\Delta$ is the order parameter, $\mu$ is the  chemical potential
(Fermi energy $\mu=\varepsilon_F$), $V({\bf{r}})$ is the potential of impurities.
We suppose here, that the magnetic field $B$ is weak
($B\ll H_{C2}$) and omit the vector potential in the eq.(2). Below we 
consider the two-dimensional case. We suppose also, that 
there is only one short range impurity (with the interaction
radius of order of $p^{-1}_F$) inside  the vortex core.

In our problem the order parameter $\Delta$ in the absence of   the impurity is given by the  expression
\begin{equation}
\Delta({\bf r})=
\Delta(r)\exp (i\varphi)
\end{equation}
where $\varphi$ is polar angle, $r=|{\bf r}|$.
The low energy excitation spectrum in the absence of the impurity was found in the
paper \cite{12}. The system (2) possesses  a  very important property:
if $E$ is an eigenvalue with the eigenfunction ($f_1,f_2$), then  $-E$
is also an eigenvalue and the corresponding eigenfunction is ($f^*_2,-f^*_1$).
This property holds in a magnetic field too.

The low energy excitation spectrum $E^0_n$ is given by the  equation \cite{12}
\begin{equation}
E^0_n=-(n-1/2)\omega_0
\end{equation}
where
\begin{displaymath}
\omega_0=\int^{\infty}_0
\frac{dr\Delta(r)}{p_Fr}e^{-2K(r)}/
\int^{\infty}_0dre^{-2K(r)} \quad,
\end{displaymath}
\begin{equation}
K(r)=\int^r_0 dr_1\Delta(r_1)/v_F \quad .
\end{equation}
If Kramer-Pesh effect takes place, then with a logarithmic
accuracy we obtain from the equations (4),(5)
\begin{equation}
\omega_0=\frac{\Delta^2}{\varepsilon_F}
\ln \left (\frac{\Delta}{T} \right)~, \quad \Delta=\Delta_{(\infty)}
\end{equation}
The eigenfunction, corresponding to the eigenvalue (4) is
\begin{equation}
\bar f_n=
\left (
\begin{array}{c}
f_1 \\
f_2
\end{array} \right)_n=
\tilde c e^{-K(r)}
\left (
\begin{array}{l}
e^{in\varphi}J_n(p_Fr) \\
-e^{i(n-1)\varphi}J_{n-1}(p_Fr)
\end{array} \right)
\end{equation}
where $\tilde c$ is the  normalization constant, $J_n(x)$ is a Bessel
function, $n=0,\pm 1, \pm 2, ...$.

For the excitation spectrum $E$ in the presence of an impurity 
inside the vortex core, we obtain from the eq.(2) the following system of equations 
\begin{equation}
\det \left ((\hat\varepsilon -E)+\hat A \right) =0
\end{equation}
where the operator $\hat A$ is given by its matrix elements. In
the basis (7) we have
\begin{equation}
A_{kn}=
< \bar f^+_k \left (
\begin{array}{ll}
V({\bf r}-{\bf a}); & 0 \\
0; & -V({\bf r}-{\bf a})
\end{array} \right) \bar f_n > \quad,
\end{equation}
{\bf a}  is the position of the impurity relative to the vortex center, and
\begin{equation}
\hat\varepsilon_{kn}=\delta_{kn}E^{0}_n ~.
\end{equation}
In the eq.(9) essential are values of $r$ such that $r   \gg p^{-1}_F$. So we can use
an asymptotic expansion of Bessel functions to find matrix elements
$A_{kn}$. A simple calculation gives
\begin{equation}
A_{kn}=e^{i(k-n)\varphi_a}
\left \{I_1(a)\cos \left (
\frac{\pi(n+k)}{2} \right) -I_2(a)\sin
\left (\frac{\pi (n+k)}{2} \right) \right \}
\end{equation}
where $\varphi_a$ is the polar angle of the vector ${\bf a}$, and the
quantities $I_{1,2}$ are given by the equation
\begin{equation}
\left (
\begin{array}{l}
I_1 \\
I_2
\end{array} \right) =
\frac{C^2}{a} e^{-2K(a)}
\int d^2rV(r)  \left (
\begin{array}{l}
\sin (2p_F(a+\frac{({\bf  a} {\bf  r}}{a})) \\
\cos (2p_F(a+\frac{({\bf  a} {\bf r})}{a}))
\end{array} \right)
\end{equation}
In the eq.(12) the  normalization constant $C$ is equal to
\begin{equation}
C^2=\{2\pi\int^{\infty}_0 dr e^{-2K(r)} \}^{-1}
\end{equation}
If there  are several impurities inside the vortex core, then the operator
$\hat A$ in the eq.(8) is a simple sum $\hat A_i$ over all impurities.
Therefore
\begin{equation}
A^{\{a_i\}}_{kn}=\sum_i A_{kn}(a_i)
\end{equation}
where $A_{kn}(a_i)$ is given by the eq.(11). It follows from
the equations (11),(14), that the transition matrix elements $A_{kn}$ are
separable. That is $A_{kn}$ can be presented as a finite sum of terms of
the type  $\tilde A^j_k \tilde B^j_n$
\begin{equation}
A_{kn}=\sum_j\tilde A^j_k\tilde B^j_n
\end{equation}
As a result  we can obtain an expression for the excitation spectrum in
an explicit form.
If only one  impurity is placed inside the vortex core, then we obtain
from the equations (8), (11), (15) the following  equation for the excitation spectrum
\begin{equation}
\det \left (
\begin{array}{ll}
1+I_1\sum \limits^{N+1}_{L=-N}
\frac{1}{\varepsilon_{2L}+E};
I_2\sum \limits^{N+1}_{L=-N}
\frac{1}{\varepsilon_{2L}+E} \\
I_2\sum \limits^{N+1}_{L=-N}
\frac{1}{\varepsilon_{2L-1}+E};
1-I_1\sum \limits^{N+1}_{L=-N}
\frac{1}{\varepsilon_{2L-1}+E}
\end{array} \right) =0
\end{equation}
For the linear spectrum given by the eq.(4), we obtain in the limit
$N\to +\infty$
\begin{eqnarray}
\nonumber
\sum \limits^{N+1}_{L=-N}
\frac{1}{\varepsilon_{2L}+E}=
\frac{\pi}{2\omega_0}\mbox{ctg}
\left (
\pi \left (\frac{1}{4}+\frac{E}{2\omega_0}
\right) \right)
\end{eqnarray}
\begin{eqnarray}
\sum \limits^{N+1}_{L=-N}
\frac{1}{\varepsilon_{2L-1}+E}=
-\frac{\pi}{2\omega_0}\mbox{ctg}
\left (
\pi \left (\frac{1}{4}-\frac{E}{2\omega_0}
\right) \right)
\end{eqnarray}
With a help of the eq.(17) we reduce the eq.(16) to the form
\begin{equation}
1+
\frac{\pi^2 \left ((I_1)^2+(I_2)^2 \right)}{4\omega^2_0}+
\frac{\pi I_1}{\omega_0\cos (\frac{\pi E}{\omega_0})}=0
\end{equation}
It follows from the eq.(18), that the low energy excitation spectrum is
strong correlated even in the presence of an impurity inside the vortex core.
If $E_0$  is a spectrum point, that is if $E_0$ is some solution
of the equation for the spectrum (18), then all solutions of the eq.(18)
are given by the equation 
\begin{equation}
E=\pm E_0+2\omega_0N~, \quad N=0, \pm 1, \pm 2~.
\end{equation}

Hence the discrete spectrum is given by two sets of equidistant points.

Functions $I_{1,2}$ are periodic with the period $\pi/p_F$. Both are
defined by the same  function with shift  by a quarter of the period.
The amplitude $I_{1.am}$ of these functions is a smooth function of the
parameter ($a/\xi$) and is given by the eq.(12).

With the accuracy of $\omega_0(\Delta/\varepsilon_F)$ a point $a_0$
exists such that
\begin{equation}
\label{20}
I_{1.am}(a_0)=
-\frac{2\omega_0}{\pi}~; \quad I_2(a_0)=0~.
\end{equation}
Hence at the points $a_0+\delta a$ given by equation
\begin{equation}
\delta a= \left (
\frac{\pi}{p_F} \right) N~, \quad N=0,\pm1, \pm2...
\end{equation}
we have
\begin{equation}
E_0=\frac{\delta a}{2} \left (\frac{\partial I_{1.am}}
{\partial a} \right)_{a_0}
\end{equation}
Equation (22) means, 
that in a vicinity of the points of the trajectory of the vortex,
given by the eq. (20), there is a set of points, separated by the distance 
$\delta a$, where spectrum lines are practically crossing (see Fig.1.
In Fig.1 the quantity $\delta a = a - a_0$ is shown, where $a_0$ is
given by the eq.(\ref{20}).)   
The vicinity of such points we denote as  {\it lagoon}. If the impact parameter 
of the trajectory is smaller than some critical value, then on such a
trajectory there are two lagoons.
When the vortex moves through these two lagoons many
excitations are created inside the vortex core. The contribution 
of these excitations to
the dissipative part of conductivity will be found below.

The situation of having several impurities inside the vortex core is considered
in Appendix. We prove there, that the strong correlation in the level positions survives
for two impurities inside the vortex core. We can make a conjecture that the  strong correlation
in level positions exists for many impurities inside the vortex core, while the condition
$l_{tr} \gg \xi $ if fulfilled. But on a  vortex trajectory of a general status lagoons with 
large number of practically crossing energy level lines do not exist, if there are two or
more impurities inside the vortex core.
\section{One impurity at small distances
($a \ll \xi(\Delta/\varepsilon_F)^{1/2})$ from the  vortex center}

First of all we shall consider one impurity with a short range
(of order of $p^{-1}_F$) potential placed exactly at the vortex center.

At  distances $\rho\gg p^{-1}_F$ from the vortex center we can use for the 
solution of the eq.(2) the quasiclassical approximation with the first order
correction terms. Indeed  these correction terms will give an expression
for spectrum. And at  small distances of order of $p^{-1}_F$ we can
omit nondiagonal elements in the eq.(2). As a result we obtain the following  
expression for the spectrum
\begin{equation}
\label{36}
E_n=-(n-1/2)\omega_0+
v_F\mbox{tg} \left (
\frac{\theta_{n-1}-\theta_n}{2} \right)
/ \int^{\infty}_0 d\rho e^{-2K(\rho)}
\end{equation}
where $\theta_n$ is the scattering phase in a state with angular
momentum $n$ in the presence of the impurity potential $V({\bf r})$.
The corresponding eigenfunction is given by the expression
\begin{equation}
\label{37}
{\bf f}_n=\left (
\begin{array}{l}
f_1 \\
f_2
\end{array}
\right)_n =
\frac{\tilde C}{2} e^{-K(\rho)} \left (
\begin{array}{l}
e^{in\varphi}(J_n(p_F\rho+\theta_n)+J_n(p_F\rho+\theta_{n-1}) \\
-e^{i(n-1)\varphi}(J_{n-1}(p_F\rho+\theta_n)+J_{n-1}(p_F\rho+\theta_{n-1}))
\end{array}
\right)
\end{equation}
where $\tilde C$ is a normalization constant.
Suppose now, that the impurity is placed on a distance ${\bf a}$
from vortex center, such that
$|{\bf a}|\ll \xi (\Delta/\varepsilon_F)^{1/2}$. In the equation (2) we
make a transformation to the coordinate system with the origin at the  impurity. Then we obtain
\begin{equation}
\label{38}
\left (
\begin{array}{ll}
-\frac{1}{2m} \frac{\partial^2}{\partial {\bf r}^2}-\mu+V({\bf r})-E; &
|\Delta|e^{i\varphi}+\left ({\bf a}\frac{\partial}{\partial {\bf r}} \right)
(|\Delta |e^{i\varphi}) \\
|\Delta |e^{-i\varphi}+\left ({\bf a}\frac{\partial}{\partial {\bf  r}} \right)
(|\Delta |e^{-i\varphi}); &
\frac{1}{2m} \frac{\partial^2}{\partial {\bf r}^2}+\mu-V(\bf{r})-E
\end{array}
\right)
\left (
\begin{array}{l}
f_1 \\
f_2
\end{array}
\right) =0
\end{equation}
From the eq.(\ref{38}) we obtain the following  equation for the excitation spectrum
\begin{equation}
\label{39}
\det \left ((\hat\varepsilon -E)+\hat A \right) =0
\end{equation}
where $\hat\varepsilon_{kn}=E_n\delta_{nk}$ and $E_n$ is given
by the eq.(\ref{36}). The operator $\hat A$ is given by matrix elements
$\hat A_{kn}$ in the basis defined in the eq.(\ref{37}).
\begin{equation}
\label{40}
\hat A_{kn}=< \bar f^+_k \left (
\begin{array}{ll}
0; & ({\bf a}\frac{\partial}{\partial {\bf r}})(|\Delta |e^{i\varphi}) \\
({\bf a}\frac{\partial}{\partial {\bf  r}})(|\Delta |e^{-i\varphi}); &  0
\end{array}
\right) {\bf f}_n >
\end{equation}
A simple straightforward calculation making use of the equations (\ref{37}), (\ref{40}) gives
\begin{displaymath}
\tilde A_{kn}=
-\pi aC^2 \int^{\infty}_0 d\rho
\frac{|\Delta_{(\rho)}|}{\rho}e^{-2K(\rho)}
\left \{
\delta_{k,n+1}e^{-i\varphi_a}
\cos \left (\frac{\theta_{n-1}-\theta_{n+1}}{2} \right) + \right.
\end{displaymath}
\begin{equation}
\label{41}
\delta_{k,n-1}e^{i\varphi_a}\cos \left (
\frac{\theta_n-\theta_{n-2}}{2} \right) \left. \right \}
\end{equation}
In the eq.(\ref{41}) the constant $C$ is given by the eq.(13). It  follows from the eq.(\ref{41}),
that in the operator in the eq.(\ref{39}), only diagonal and near diagonal
elements are nonzero. Now we define the function $B(I,E,n)$ in the following  manner
\begin{equation}
\label{42}
B(I,E,n-1)=
-E_n-E-\frac{I^2\cos^2 \left (\frac{\theta_{n-1}-\theta_{n+1}}{2}
\right)}{B(I,E,n)} \quad,
\end{equation}
where
\begin{displaymath}
 I=\pi aC^2 \int^{\infty}_0 d\rho
\frac{|\Delta |}{\rho}\cdot e^{-2K(\rho)}, \quad
I/\omega_0=p_Fa/2.
\end{displaymath}
With the help of the function $B$
we reduce  the eq.(\ref{39}) for the spectrum to the following  simple form
\begin{equation}
\label{43}
\det \left (
\begin{array}{lll}
B(I,E,1);  -Ie^{i\varphi_a}\cos \left (
\frac{\theta_0-\theta_2}{2} \right);  0; 0 \\
-I e^{-i\varphi a}\cos \left (\frac{\theta_0-\theta_2}{2}
\right);  E_0-E;  -Ie^{i\varphi_a};  0 \\
0;  -I e^{-i\varphi_a};  E_1-E;
-Ie^{i\varphi_a}\cos \left (\frac{\theta_0-\theta_2}{2} \right) \\
0;  0;  I e^{-i\varphi_a}\cos \left (
\frac{\theta_0-\theta_2}{2} \right);  B(I,-E,1)
\end{array} \right) =0
\end{equation}
The equation (\ref{42}) means, that $B(I,E,1)$ can be presented as an infinite fraction
\begin{equation}
\label{44}
B(I,E,1)=-E_2-E-
\frac{I^2\cos^2 \left ((\theta_1-\theta_3)/2 \right)}
{\displaystyle -E_3-E-\frac{I^2\cos^2((\theta_2-\theta_4)/2)}
{\displaystyle -E_4-E-\frac{I^2\cos^2((\theta_3-\theta_5)/2)}
{\displaystyle ... \; B(I,E,n+1)}}}
\end{equation}
The fraction (\ref{44}) converges very quickly, if for $B(I,E,n+1)$ we use the
expression
\begin{equation}
\label{45}
B(I,E,n+1)=
\left [(n+1/2) \omega_0-E+
\sqrt{((n+1/2)\omega_0-E)^2-4I^2} \right] /2~,
\end{equation}
\begin{equation}
\label{46}
(n+1/2)\omega_0\pm E \gg |I|
\end{equation}
Suppose now, that the impurity is of a small size, so only $S$-scattering is
essential. Suppose also, that the
impurity potential is of order of the atomic one and hence the inequality takes place
\begin{equation}
\label{47}
\varepsilon_0=\Delta\mbox{tg}(\theta_0/2) \gg \omega_0
\end{equation}
Then in the first approximation the equation (\ref{43}) for low energy excitations uncouples 
to two independent branches
\begin{equation}
\label{48}
B(I,E,1)=0 \quad \mbox{and} \quad B(I,-E,1)=0
\end{equation}

And hence we obtain two independent families of spectrum lines.
Of course in this approximation they will cross. And only in
the next  approximation with respect to   the parameter $(\omega_0/\varepsilon_0)^2$
a gap in the crossing points will open.

For small values of $E$ we have
\begin{equation}
\label{49}
B(I,E,1)=B+\alpha E
\end{equation}
where
\begin{equation}
\label{50}
\alpha = \frac{\partial B(I,E,1)}{\partial E}
\end{equation}
Inserting the expression (\ref{49}) into the eq.(\ref{43}) we obtain the following  equation for the
lowest energy level near the crossing points
\begin{equation}
\label{51}
E^2\alpha^2\varepsilon^2_0=
B^2(I\cos (\theta_0/2))^2+
(\varepsilon_0B+(I\cos (\theta_0/2))^2)^2
\end{equation}
Hence the value of the gap $\delta$ near this crossing point is
equal to
\begin{equation}
\label{52}  
\delta=
\frac{|I\cos (\theta_0/2)|^3}
{| \alpha\varepsilon_0| \sqrt{\varepsilon^2_0+
(I\cos (\theta_0/2))^2}}
\end{equation}
In the  eq.(\ref{51}) the quantity $\alpha$ should be taken at  the point
\begin{equation}
\label{53}
\alpha=
\frac{\partial B(I,E,1)}{\partial E} \left |_{B=-
\frac {\varepsilon_0(I\cos (\theta_0 / 2))^2}
{(I\cos (\theta_0 / 2))^2+\varepsilon^2_0}} \right.
\end{equation}
By  the order of magnitude the value of the gap $\delta$ is
given by the equation
\begin{equation}
\label{54}
\delta\sim\Delta (a/\xi)^3
\end{equation}

It is possible to keep in the expansion of the order parameter $\Delta$ with respect to the
shift ${\bf a}$ in the eq.(\ref{38}) terms up to the second order in  ${\bf  a}$. Then
the
operator $\hat A$ will have the following  nonzero matrix elements:
$A_{nn}; A_{n,n\pm 1}; A_{n,n\pm 2}$. The equation (\ref{39}) in this
approach enables us to determine the excitation spectrum up to the
shift ${\bf a}$ of  order of
\begin{equation}
\label{55}
a\sim\xi (\Delta/\varepsilon_F)^{1/3}
\end{equation}
At the boundary of this region the gap $\delta$, given by the eq.(\ref{54}), is of  order
of $\omega_0$. Hence energy levels can "cross" only in the lagoons (eq.(22)),
or when the impurity is placed near the vortex core (eq.(\ref{52})).

The equation (\ref{43}) for the  spectrum can be reduced to the form
\begin{eqnarray}
\label{55a}
\nonumber
\left(\frac {I^2} {E_0} \ cos^2 \left (\frac{\theta_0 - \theta_2}{2} \right) + 
\frac{B(I,E,1)+B(I,-E,1)}{2} \right)^2- \\  
\nonumber
\left(\frac{B(I,E,1)-B(I,-E,1)}{2}- 
\frac{EI^2}{E^2_0}\cos^2 \left (\frac{\theta_0-\theta_2}{2}
\right) \right)^2+ \\
\frac{I^2-E^2}{E^2_0}
B(I,E,1)B(I,-E,1)+
\left (\frac{EI^2}{E^2_0}\cos^2 \left (
\frac{\theta_0-\theta_2}{2} \right) \right)^2=0 \quad\quad\quad
\end{eqnarray}

Near the crossing points the last two terms are small (of order of
$(I^3/E_0)^2~)$ and lead to the repulsion of spectrum lines. In
zero approximation neglecting two last terms the 
eq.(\ref{55a}) uncouples   to give two families of independent spectrum lines.

Energy levels as a function of the shift {\bf a} (or quantity $I\sim a$) are given
in the  Fig. 2. Inside the circle in the Fig. 2 the equations for two spectrum
branches are 

\begin{eqnarray}
E / \omega_0 -1.073 = \delta (0.1335 t \pm   \sqrt{1+t^2}); \\
t = 1.343 (I / \omega_0 -2.3172) / \delta; \\
\delta = 3.2 \cdot 10^{-3}
\end{eqnarray}

The contribution of these "crossing" points to the
dissipative part of conductivity will be discussed below.

\section{Landau-Zener tunnelling near the crossing point of
spectrum lines}

Near the crossing point of spectrum lines, given by the eq.(18), we
can put
\begin{equation}
\label{56}
\frac{\pi I_1}{2\omega_0}=-1+y; \quad
\frac{\pi I_2}{2\omega_0}=2p_FX; \quad
X=Vt
\end{equation}
Where $V$ is the velocity of the vortex. For two close spectrum points
$E_{\pm}$ we obtain from the eq.(18) the following  value
\begin{equation}
\label{57}
E_{\pm}=
\pm \varepsilon_{(t)}\frac{\omega_0}{\pi}~, \quad
\varepsilon_{(t)}=\sqrt{y^2+(2p_FX)^2}
\end{equation}

The usual Landau-Zener consideration leads to the following value for  the probability
$W_{++}$ for a "particle" to remain on the same branch after
collision
\begin{equation}
\label{58} 
W_{++}=1-\exp \left (-\frac{\omega_0y^2}{2p_F|V|} \right)~.
\end{equation}
With a help of the eq.(\ref{58}) we will  find the energy, transmitted to the vortex
at one collision with impurity.

\section{Dissipative part of conductivity in superconductors}

 Let us first consider the superclean limit, when  
only one impurity can be found inside the vortex core.
Then for small values of the vortex velocity
$V$, such that
\begin{equation}
\label{59}
V \ll v_F (\Delta /\varepsilon_F)^2~,
\end{equation}
an excitation can arise only at spectrum line crossing points. Such
crossing points are located only in lagoons (eq.(20), (21)), or
if the vortex center is close to the impurity (eq. (\ref{52})).

Suppose, that between two consequent collisions the system goes to the equilibrium due
to inelastic scattering processes. In such a case the  dissipative part of
conductivity is directly connected to the  energy, stored by vortex
after one collision with impurity.

Consider first processes happening in lagoons. Suppose, that $\rho_0$ is
the impact parameter. Then near the point $a_0$ (eq. (20)) we have
\begin{equation}
\label{60}
\frac{\pi I_{1.am}}{2\omega_0}=1+y~, \quad
y=\frac{\pi}{2\omega_0} \left (
\frac{\partial I_{1.am}}{\partial\rho} \right)_{a_0}
x\sqrt{1-(\rho_0/a_0)^2}
\end{equation}
Periodicity $\delta x$ of a crossing point with respect to $x$ is
\begin{equation}
\label{61} 
\delta x=\frac{\pi}{p_F \sqrt{1-(\rho_0/a_0)^2}}
\end{equation}
Hence the  full number $N$ of crossing points in a lagoon, with effective
transition of "particle" to the other branch, is
\begin{equation}
\label{62}
N=\frac{2x}{\delta x}=
\frac{4 \sqrt{2}}{\pi^2}
\frac{p_F(p_FV\omega_0)^{1/2}}
{\left (\frac{\partial I_{1.am}}{\partial\rho} \right)_{a_0}}
\end{equation}
To obtain the eq.(\ref{62}) we use the eq.(\ref{58}). We get with a help of the eq.(\ref{62}), that
the vortex, after passing through two lagoons (one collision with impurity),
stores the energy $\delta E$ being equal to
\begin{equation}
\label{63}
\delta E=4\omega_0N^2=
\frac{128}{\pi^4}
\frac{\omega^2_0p^3_FV}
{(\partial I_{1.am/\partial\rho})^2_{a_0}}
\end{equation}
As a result, we can estimate the contribution of transitions in lagoons
to the dissipative part of conductivity $\sigma^{(1)}_{diss}$
\begin{equation}
\label{64}
\sigma^{(1)}_{diss}=\frac{256}{\pi^4}
\frac{a_0n_{imp}}{B\phi_0}
\frac{\omega^2_0p^3_F}{(\partial I_{1.am/\partial\rho})^2_{a_0}}
\end{equation}
where $\phi_0=\pi/e$ - flux quantum, $B$ - magnetic field value.
Consider now the energy, dissipated as an impurity passes near the vortex core.
From the eq.(\ref{43}), (\ref{52}) we obtain the following  expression for the excitation spectrum
near the crossing points
\begin{equation}
\label{65}
E=\pm \left (
\frac{I^6}{\varepsilon^4_0\gamma^2_2} +
(\omega_0p_F(\delta a)/2)^2 \right)^{1/2}
\end{equation}
where
\begin{equation}
\label{66}
\gamma_2=
\frac{\partial B(I,E,1)}{\partial I}~; \quad
I=\omega_0p_Fa/2
\end{equation}

$(\delta a)$ is a shift from a "crossing point",
the parameter $\gamma_2$ is of order of unity. According to the
eq.(\ref{58}), the probability $W_{++}$ for the particle to remain on the same
branch after  a collision is equal to
\begin{equation}
\label{67}
W_{++}=1-\exp \left \{
-\frac{\pi a^6(\omega_0p_F/2)^5}
{\varepsilon^4_0\gamma^2_2|V|} \right \}
\end{equation}

Hence the number of excitations $N$, that arise in the vortex core, when the
impurity passes through the vortex near its center, is
\begin{equation}
\label{68}
N=\frac{2\tilde a p_F}{\pi}~,  \quad
\tilde a = \left (
\frac{V\varepsilon^4_0\gamma^2_2}
{\pi (\omega_0 p_F/2)^5} \right)^{1/6}
\end{equation}

These transitions give the following  contribution to the energy
dissipation $\delta E$ per volume and time unit
\begin{equation}
\label{69}
\delta E=2\tilde a V\omega_0N^2n_{imp}(B/\phi_0)
\end{equation}
Equations (\ref{64}), (\ref{69}) completely determine the dissipative part of
the current
\begin{equation}
\label{70}
j_{diss}=\sigma^{(1)}_{diss} E+
\frac{8p^2_F}{\pi^2} \cdot
\frac{\omega_0n_{imp}}{\phi_0}
\left (
\frac{\varepsilon^4_0\gamma^2_2}
{\pi(\omega_0p_F/2)^5}
\right)^{1/2}(E/B)^{1/2}
\end{equation}
By the order of magnitude we have
\begin{equation}
\label{71}
\left (
\frac{\partial I_{1.am}}{\partial\rho} \right)_{a_0}
\sim \frac{2\omega_0}{\pi\xi}
\end{equation}
Hence, in the range of velocities $V$, such that
\begin{equation}
\label{72}
V/v_F>(\Delta/\varepsilon_F)^4
\end{equation}
the second term in the eq.(\ref{70}) is smaller than the first one. By the
large parameter
\begin{equation}
\label{73}
\frac{(n_{im}\xi^2)}{\tau_{tr}\Delta}(\varepsilon_F/\Delta)^2
\end{equation}
the dissipative part of conductivity, given in the eq.(\ref{70}), exceeds the
quasiclassical value for conductivity in the two dimensional case.
The expression (\ref{64}) for conductivity was obtained under the
assumption, that the number of crossing point in a lagoon eq.(\ref{62})
is large. This condition gives the same restriction on the
value of the velocity as  given in the eq.(\ref{72}).

\section{Conclusion}

In the two dimensional case the excitation spectrum in vortex core is
discrete. This results the strong increase of the dissipative
part of conductivity $\sigma_{xx}$ (eq.(1)) comparing to 
its value obtained by the quasiclassical method. It is very probable,
that the strong increase of conductivity $\sigma_{xx}$ at low
temperatures, obtained in experimental papers \cite{1,2}, is
related  to the phenomena, considered in this paper. For the
detailed comparison to the experiment both experimental and
theoretical investigations are necessary.

Consider now the applicability region of the equations. (\ref{64}), (\ref{70}), (\ref{73}).
We do not see any restriction for the temperature $T$ in
the framework of the  model used except $T\ll T_c$. In the range
$T\sim T_c$ excitations with the energy $\varepsilon\sim\Delta$ are
essential. In this region the excitation spectrum is not equidistant.
Probably its more essential, that in our model the main mechanism
of dissipation is the scattering of excitations off  impurities.
In real high-$T_c$ compounds such an assumption holds only in the
low temperature region. The predicted effect should vanish
in the strong magnetic field region $H\sim H_{c2}$.

More complicated are restrictions for the electrical field magnitude  or for 
the vortex velocity. The restriction $v_F(\Delta/\varepsilon_F)^3\ll
V\ll v_F(\Delta/\varepsilon_F)^2$, given in the introduction,   holds 
in  our approximation. For the  vortex velocity
$V\gg v_F(\Delta/\varepsilon_F)^2$ the adiabatic consideration is
inapplicable. An impurity on the distances $a<\xi$ from
vortex center leads to the transitions of quasiparticles to 
highly exited states. In the region $V\ll v_F(\Delta/\varepsilon_F)^3$
the small gap in the excitation spectrum of order of
$\omega_0(\Delta/\varepsilon_F)^3$ leads to the adiabaticity of motion
in  lagoons. In this case
the impurities, that pass at small distances of order of
$\xi(\Delta/\varepsilon_F)^{1/2}$ from the vortex center (second term in
the eq.(
\ref{70})) give the main contribution to the energy
dissipation. It is essential, that even in the frame  of our approach
the $I-V$ characteristic changes its form for different ranges  of 
the velocity.

There are also some other  mechanisms, that can change the $I-V$
characteristic. At small values of vortex velocities pinning can
start to be essential. It can reduce dissipation. The pinning force
is strongly  dependent on the  interaction between vortices and hence
on the magnetic field magnitude  and anisotropy factor. In superclean
superconductors pinning can be considered  weak.

The tunnelling of excitations to the neighbouring planes can reduce
dissipation at large velocity. If the anisotropy factor $\varepsilon$
is small then the spectrum near the crossing points does not change much and the 
Landau-Zener tunnelling probability does not change. But near the
points, where the energy of a quasiparticle is close to the unperturbed
position of a level from  the neighbouring plane,  a notable probability arises
for tunnelling of a quasiparticle to the neighbouring plane. As
a result, the quasiparticle energy does not  increase any more and
dissipation will be smaller than given by the eq.(\ref{63}). The energy
relaxation of quasiparticles can change $I-V$ characteristic as
for large values of vortex velocity so at small one. If the energy
relaxation time is large enough, then effective heating of excitations
inside of the vortex core takes place. The effective temperature reaches a  
value of order of $T_c$ \cite{15}. The value of $\sigma_{xx}$ decreases
in this case. On the other hand, strong energy relaxation can lead
to the equilibrium inside of one act of tunnelling. This phenomenon
will decrease $\sigma_{xx}$ value in the range of small vortex
velocities. Now the physical reason of the energy relaxation in high-$T_c$
superconductors at low temperatures is not clear. And we can not  make any
quantitative estimation of this effect. Equations (\ref{64}), (\ref{70}) are
valid in the superclean limit, if at  distances of order of $\xi$ from the
vortex center there is  no more than one impurity. If at a 
distance of order of $\xi$ from the vortex center there are two impurities,
then the probability of a level crossing decreases strongly. As a  result,
the value of $\sigma_{xx}$ decreases too.

$\sigma_{xx}$ as a function of impurity
concentration has a maximum at $\omega_0\tau_{tr}$ of order of
one. The question about the value of the impurity concentration,
for which the  matrix can be considered as random demands an additional
investigation.

\acknowledgments

The research of Yu.N.Ovchinnikov supported by the CRDF grant
RP1-194.

\appendix
\section{A large number  of impurities inside of the vortex core}

If inside the  vortex core a large number  of impurities is placed, then
transition matrix elements $A_{kn}\{\bf{a_i}\}$ ($\bf{a_i}$ is the  position of
$i$-th impurity), are  given by the eq.(14). Equation for the spectrum in 
such a case is
\begin{equation}
\label{23}
\det (\hat 1+\hat C)=0
\end{equation}
with the matrix elements of the operator $\hat C$ being equal to
\begin{equation}
\label{24}
\hat C_{jj_1}=
\sum_k \frac{\tilde B^j_k\tilde A_k^{j_1}}
{E_k-E}~, \quad \hat 1_{nm}=\delta_{nm}
\end{equation}
If $M$ impurities are placed inside the vortex core, then the size of the matrix
in the eq.(\ref{23}) is (4Mx4M). Due to symmetry properties of the
elements $A_{kn}$ (eq.(11)), this matrix can be easily reduced to the
size (2Mx2M). The structure of this matrix $\hat M$ is simple. On
the diagonal are placed blocks (2x2), defined only by one impurity.
The second type of blocks  are blocks (2x2), that give the interference
contribution of a pair of impurities ($ij$).

To clarify the structure of the matrix $\hat M$ in the general case, we
give below the explicit expression of the matrix $\hat M$ for two impurities
inside the vortex core.
\begin{equation}
\label{25} 
\hat M= \left (
\begin{array}{lll}
1 + \frac{\pi I^1_1}{2\omega_0} \mbox{ctg}
\left (\frac{\pi}{4}+\frac{\pi E}{2\omega_0} \right);
\frac{\pi I^1_2}{2\omega_0} \mbox{ctg} \left (\frac{\pi}{4}+
\frac{\pi E}{2\omega_0} \right);
\frac{I^2_1}{2\omega_0}z_1~;   \frac{I^2_2}{2\omega_0}z_1 \\
-\frac{\pi I^1_2}{2\omega_0} \mbox{ctg} \left (\frac{\pi}{4}-
\frac{\pi E}{2\omega_0} \right);
1 + \frac{\pi I^1_1}{2\omega_0} \mbox{ctg}
\left (\frac{\pi}{4}-\frac {\pi E}{2\omega_0} \right);
\frac{I^2_2}{2\omega_0}z;
-\frac{I^2_1}{2\omega_0}z \\
\frac{I^1_1}{2\omega_0}z^*_1;
\frac{I^1_2}{2\omega_0}z^*_1;
1+\frac{\pi I^2_1}{2\omega_0} \mbox{ctg} \left (\frac{\pi}{4}+
\frac{\pi E}{2\omega_0} \right);
\frac{\pi I^2_2}{2\omega_0} \mbox{ctg} \left (
\frac{\pi}{4}+
\frac{\pi E}{2\omega_0} \right) \\
\frac{I^1_2}{2\omega_0}z^*;
-\frac{I^1_1}{2\omega_0}z^*;
-\frac{\pi I^2_2}{2\omega_0} \mbox{ctg}
\left (\frac{\pi}{4}-\frac{\pi E}{2\omega_0} \right);
1 + \frac{\pi I^2_1}{2\omega_0} \mbox{ctg}
\left (\frac{\pi}{4}-\frac{\pi E}{2\omega_0} \right)
\end{array}
\right)
\end{equation}
In the eq.(\ref{25}) the upper index in  $I^j_i$ means the number of
impurities ($j=1,...M$) and  the lower index $i$ can be 1 or 2. The quantities
$z,z_1$ are defined by the equations
\begin{displaymath}
z=\sum^{\infty}_{L=-\infty}
\frac {e^{2iL(\varphi_{a_1}-\varphi_{a_2})}}
{L-1/4+E/2\omega_0}
\end{displaymath}
\begin{equation}
\label{26}
z_1=\sum^{\infty}_{L=-\infty}
\frac {e^{i(2L+1)(\varphi_{a_1}-\varphi_{a_2})}}
{L+1/4+E/2\omega_0}
\end{equation}
From the eq.(\ref{26}) it follows, that
\begin{equation}
\label{27}
z_1(E)=-z^*(-E)e^{i(\varphi_{a_1}-\varphi_{a_2})}
\end{equation}
A straightforward calculation gives for the quantity $z(E)$ the following  
expression
\begin{equation}
\label{28}
z(E)=
-\frac {\pi}{\sin \left (\frac {\pi}{4}-
\frac {\pi E}{2\omega_0} \right)}
e^{-i\pi(1/4-E/2\omega_0)+
iX_{a_1a_2}(1/2-E/\omega_0)}
\end{equation}
where
\begin{equation}
\label{29}
X_{a_1a_2}=(\varphi_{a_1}-\varphi_{a_2})/\mbox{mod} \; \pi > 0
\end{equation}
From the eq.(\ref{25}) we obtain
\begin{eqnarray}
\nonumber
\det\hat M= \left (
1+J_1+
\frac{\pi I^1_1}{\omega_0\cos \left (\frac{\pi E}{\omega_0} \right)}
\right)
\left (1+J_2+\frac{\pi I^2_1}{\omega_0\cos \left (
\frac{\pi E}{\omega_0} \right)}
\right) +
\end{eqnarray}
\begin{eqnarray}
\nonumber
\frac{I^1_2I^2_2}{4\omega^2_0}(z_1z^*+zz^*_1)+
\frac{|zz_1|^2}{16\omega^4_0}J_1J_2-
\end{eqnarray}
\begin{eqnarray}
\nonumber
\frac{|z|^2}{4\omega^2_0} \left [I^1_1+
\frac{2\omega_0}{\pi}J_1 \mbox{ctg} \left (\frac{\pi}{4}+
\frac{\pi E}{2\omega_0} \right) \right]
\left [I^2_1+\frac{2\omega_0}{\pi}J_2 \mbox{ctg} \left (
\frac{\pi}{4}+\frac{\pi E}{2\omega_0}
\right) \right] -
\end{eqnarray}
\begin{eqnarray}
\label{30}
\frac{|z_1|^2}{4\omega^2_0} \left [I^1_1+
\frac{2\omega_0}{\pi}J_1 \mbox{ctg} \left (\frac{\pi}{4}-
\frac{\pi E}{2\omega_0} \right) \right]
\left [I^2_1+\frac{2\omega_0}{\pi}J_2 \mbox{ctg} \left (
\frac{\pi}{4}-\frac{\pi E}{2\omega_0}
\right) \right]
\end{eqnarray}
where
\begin{equation}
\label{31}
J_i=\frac {\pi^2}{4\omega^2_0}
\left ((I^i_1)^2+(I^i_2)^2 \right )
\end{equation}

Note, that the quantity $zz^*_1$ is pure imaginary. Hence the only term in the
eq.(\ref{30}) that has a periodical dependence with respect to the energy $E$ with the period,
differing  from $2\omega_0$, drops out. As a result, the expression
for the spectrum (19)
holds also in the case, if there are two impurities inside the vortex. Our conjecture
is: for the low energy excitation spectrum eq.(19) is correct in the clean limit
$(\tau\Delta\gg 1)$, even if there are a lot of impurities inside the vortex core.

With a help of the equations (\ref {27}), (\ref {28}), (\ref{30}) we obtain the following  equation for  
the excitation
spectrum
\begin{displaymath}
\det \hat M=0,
\end{displaymath}
\begin{displaymath}
\det \hat M=
\left (1+J_1+
\frac{\pi I^1_1}{\omega_0\cos \left (\frac{\pi E}{\omega_0} \right)}
\right)
\left (1+J_2+
\frac{\pi I^2_1}{\omega_0\cos \left (\frac{\pi E}{\omega_0} \right)}
\right)
+ \frac{4}{\cos \left (\frac{\pi E}{\omega_0} \right)} J_1J_2 -
\end{displaymath}
\begin{displaymath}
\frac{\pi^2}{4\omega^2_0\sin^2 \left (
\frac{\pi}{4}-\frac{\pi E}{2\omega_0} \right)}
\left (I^1_1+\frac{2\omega_0}{\pi}J_1
\mbox{ctg} \left (
\frac{\pi}{4}+\frac{\pi E}{2\omega_0} \right) \right)
\left (I^2_1+\frac{2\omega_0}{\pi}J_2\mbox{ctg}
\left (\frac{\pi}{4}+\frac{\pi E}{2\omega_0} \right) \right) -
\end{displaymath}
\begin{equation}
\label{32}
\frac{\pi^2}{4\omega^2_0\sin^2 \left (
\frac{\pi}{4}+\frac{\pi E}{2\omega_0} \right)} \cdot
\left (I^1_1+\frac{2\omega_0}{\pi}J_1
\mbox{ctg} \left (
\frac{\pi}{4}-\frac{\pi E}{2\omega_0} \right) \right)
\left (I^2_1+\frac{2\omega_0}{\pi}J_2\mbox{ctg}
\left (\frac{\pi}{4}-\frac{\pi E}{2\omega_0} \right) \right)
\end{equation}
For the $I-V$ characteristic in the mixed state decisive is the value (or the
existence) of the gap in the  excitation spectrum. For this reason we
will calculate the value of $\det\hat M$ for $E=0$. Simple
calculations with a help of the eq.(\ref{32}) give
\begin{equation}
\label{33}
\det\hat M(E=0)=
\frac{\pi I^2_1}{\omega_0}(1-J_1)+
\frac{\pi I^1_1}{\omega_0}(1-J_2)+
(1+J_1)(1+J_2)
\end{equation}
The expression (\ref{33}) is nonnegative. It can be written as a sum of nonnegative
terms
\begin{displaymath}
\det\hat M(E=0)=
\left (1+\frac{\pi I^1_1}{2\omega_0}+\frac{\pi I^2_1}{2\omega_0}-
\frac{\pi I^1_1}{2\omega_0} \frac{\pi I^2_1}{2\omega_0}
\right)^2+
\end{displaymath}
\begin{equation}
\label{34}
\left (\frac{\pi I^1_2}{2\omega_0} \right)^2
\left (\frac{\pi I^2_2}{2\omega_0} \right)^2 +
\left (\frac{\pi I^1_2}{2\omega_0} \right)^2
\left (1-\frac{\pi I^2_1}{2\omega_0} \right)^2+
\left (\frac{\pi I^2_2}{2\omega_0} \right)^2
\left (1-\frac{\pi I^1_1}{2\omega_0} \right)^2
\end{equation}

It means, that on the vortex trajectory of a general status,
lagoons with large number of practically crossing energy levels lines
do not exist. Now we can formulate the following  conjecture. In the clean limit
$(\tau\Delta\gg 1)$, if several impurities are placed inside  the vortex core
, the quantity $\det\hat M(E=0)$ is a sum of nonnegative
terms of the form (\ref{34}).

Now, for small values of  $E\ll \omega_0$, we obtain from the eq.(\ref{32})
\begin{displaymath}
\det\hat M(E)=
\det\hat M(E=0)-
\left ( \frac{\pi E}{\omega_0} \right)^2 \cdot
\end{displaymath}
\begin{equation}
\label{35}
\left \{2J_1J_2-\frac{\pi I^1_1}{2\omega_0}(1-J_2)-
\frac{\pi I^2_1}{2\omega_0}(1-J_1) \right \}
\end{equation}
The equations (\ref{33}), (\ref{34}), (\ref{35}) completely determine the excitation spectrum
near the  crossing points of the energy levels.

\begin{figure}
\caption{The excitation spectrum as a function of the impurity distance from the
vortex center. The parameter $(\pi^2/2\omega_0 p_F)(\partial I_{1,am}/\partial
a\dots)$ equals 0.02. The quantity $\delta a = a - a_{0}$, with $a_{0}$ is given by the eq. (20).}
\label{fig:1}
\end{figure}

\begin{figure}
\caption{The low energy excitation spectrum at small distances $a$ from the
impurity to the vortex center; $I/\omega_0=p_Fa/2$, see eq.(29); $\theta\ll 1 $
; $E_0/\omega_0=50$.}
\label{fig:2}
\end{figure}

\begin{references}
\bibitem[\dagger]{PA} Permanent address: L.D.~Landau Institute for Theoretical
Physics Academy of Sciences of Russia. Kosigin Str. 2, 117334 Moscow, Russia.

\bibitem{1} Y. Matsuda, N.P. Ong, Y.F. Yan, J.M. Harris and J.B. Peterson.
Phys. Rev. {\bf B 49}, 4380 (1994).
\bibitem{2} S.G. Doettinger et al. Europhysics Letters
{\bf 30}, 549 (1995).
\bibitem{3} L.P. Gor'kov, N.B. Kopnin. Sov. Phys. JETP {\bf 38}, 195 (1973).
\bibitem{4} J. Bardeen, M.J. Stephen. Phys. Rev. {\bf 140}, A1197 (1965).
\bibitem{5} J. Bardeen, R.D. Sherman. Phys. Rev. {\bf B12}, 2634
(1975).
\bibitem{6} A.I. Larkin, Yu.N. Ovchinnikov. Zh. Eks. Theor. Fiz. Pis'ma,
{\bf 23}, 210 (1976).
\bibitem{7} L. Kramer, W. Pesh. Z. Phys. {\bf 269}, 59 (1974).
\bibitem{8} N.B. Kopnin, A.V.Lopatin. Phys. Rev. {\bf B51}, 15291 (1995).
\bibitem{9} N.B. Kopnin, V.E. Kravtsov. Pis'ma Zh. Eks. Theor. Fiz.
{\bf 23}, 631 (1976).
\bibitem{10_1} F. Guinea, Yu. Pogorelov Phys. Rev. Lett. {\bf 74}, 462 (1995).
\bibitem{10} M.V. Feigel'man, M.A. Skvorzov. cond.-mat/9609173.
Subm. for publ. Phys. Rev. Lett. (1997).
\bibitem{11} A. Atland and M.R. Zirnbauer, cond.mat/9602137.
\bibitem{12} C. Caroli, P.G. De Gennes, J. Matricon. Phys. Rev. Lett.
{\bf 9}, 307 (1964).
\bibitem{13} A.A. Abrikosov, L.P. Gor'kov, I.E. Dyzaloshinskii.
"The methods of Quantum Field Theory in Statistical Physics".
State Publ. Corp. of Phys.-Math. Lit.,  Moscow, 1962.
\bibitem{14} L.D. Landau, E.M. Lifshitz. "Quantum mechanics. Nonrelativistic
theory". State publish. Corp. of Phys.-Math. Lit.,  Moscow, 1963.
\bibitem{15} A.I. Larkin, Yu.N. Ovchinnikov. Zh. Eks. Theor. Phys.
{\bf 73}, 299 (1977).
\end{references}
\end{document}